\documentclass[sigconf]{acmart}

\usepackage{booktabs}
\usepackage{caption}
\usepackage{subfig}
\usepackage{url}
\usepackage{epstopdf}
\usepackage{array}
\usepackage{pifont}
\usepackage{multirow}
\usepackage{amssymb}
\usepackage{balance}
\usepackage[ruled]{algorithm2e}
\usepackage{algorithmic}

\begin{document}

\setcopyright{none}
\copyrightyear{2018}
\acmConference{ArXiV}{April}{2018}
\acmPrice{}
\acmISBN{XXX}
\acmDOI{XXX}

\title[Cross-Lingual IR in Presence of Topically Aligned Corpora]
{Learning Multilingual Embeddings for Cross-Lingual Information Retrieval in the
Presence of Topically Aligned Corpora}

\author{Mitodru Niyogi}
\affiliation{\institution{Computer Science and Engineering, Indian Institute of Technology, Kanpur, India}}
\email{mitodru@cse.iitk.ac.in}

\author{Kripabandhu Ghosh}
\affiliation{\institution{Computer Science and Engineering, Indian Institute of Technology, Kanpur, India}}
\email{kripa@cse.iitk.ac.in}

\author{Arnab Bhattacharya}
\affiliation{\institution{Computer Science and Engineering, Indian Institute of Technology, Kanpur, India}}
\email{arnabb@cse.iitk.ac.in}
\orcid{0000-0001-7331-0788}

\renewcommand{\shortauthors}{M. Niyogi, K. Ghosh, A. Bhattacharya}

\begin{abstract}
	Cross-lingual information retrieval is a challenging task in the absence of
	aligned parallel corpora. In this paper, we address this problem by
	considering topically aligned corpora designed for evaluating an IR setup.
	To emphasize, we neither use any sentence-aligned corpora or
	document-aligned corpora, nor do we use any language specific resources such
	as dictionary, thesaurus, or grammar rules.  Instead, we use an embedding
	into a common space and learn word correspondences directly from there.  We
	test our proposed approach for bilingual IR on standard FIRE datasets for
	Bangla, Hindi and English. The proposed method is superior to the
	state-of-the-art method not only for IR evaluation measures but also in
	terms of time requirements. We extend our method successfully to the
	trilingual setting.
\end{abstract}

\maketitle



\section{Introduction}
\label{sec:intro}

Cross-lingual information retrieval, where multiple languages are used
simultaneously in an information retrieval (IR) task, is an important area of
research.  The increasing amount of non-English data available through the
Internet and processed by several modern age IR/NLP (natural language
processing) tasks has magnified the importance of cross-lingual IR manifold.
In particular, we address the general ad-hoc information retrieval task where
the query is in any of the $n$ languages, and retrieval can be from any of the
remaining languages.  In countries such as India where multiple languages are
used officially and regularly by a large amount of computer-educated citizens,
the above task is particularly important, and can be a game changer for many of
the digital initiatives that governments across the world are actively
promoting.

Such queries can be quite common.  For example, in nation-wide events such as
general elections, or an emergency situation, a sports event, etc., queries
like ``How many seats have party X won in state Y?'' are quite common and will
be issued in several languages.  The proposed system should be able to retrieve
the answer from documents written in any language.

Most of the previous work on cross-lingual IR \cite{Wang2012, Hermann2014}
require \emph{sentence-aligned} parallel data and other language specific
resources such as dictionaries.  Vulic et al. \cite{Vulic2015} removed this
extremely constraining requirement and learned \emph{bilingual} word embedding
using only \emph{document-aligned} comparable corpora. However, such aligned
corpora is not always readily available and need considerable effort to be
built.  Resource-poor languages such as the Indian languages specifically
suffer from this setback.

To this end, we present a multi-lingual setup where we build a cross-lingual IR
system that requires no such aligned corpora or language specific resources. It
automatically learns cross-lingual embeddings using merely TREC-style test
collections. We also propose to build a \emph{multi-lingual} embedding on the
same setup. This eliminates the requirement of building embeddings for
collection pairs in a cross-lingual retrieval paradigm as well as the need to
train bilingual embedding for all possible language pairs.  Instead, this
single multi-lingual embedding will leverage automatic cross-lingual retrieval
between any two pairs of languages. The proposed setup is particularly useful
in online situations in multi-lingual countries such as India.

To the best of our knowledge, this is the first cross-lingual IR work that
works on more than $2$ languages directly and simultaneously without targeting
each pair separately.

Our proposed method yields considerable improvements over Vulic et al.
\cite{Vulic2015} in the bilingual setting on standard Indian language test
collections.  We further demonstrate the efficacy of our method by using a
trilingual embedding.


\section{Methodology}
\label{sec:methodology}

A traditional ad-hoc information retrieval test collection (in the binary
relevance setup) is defined as $\mathbb{C} = \{\mathbb{D}, \mathbb{Q},
\mathbb{R}\}$ where $\mathbb{D}$ is a set of documents, $\mathbb{Q}$ a set of
queries, and {\it relevance} $\mathbb{R}$ is a mapping defined as $\mathbb{R}:
\mathbb{Q} \times \mathbb{D} \rightarrow \{0,1\}$.  For each document $d \in
\mathbb{D}$ and query $q \in \mathbb{Q}$, $\mathbb{R}(d,q) = 1$ if $d$ is
relevant for $q$, and $0$ otherwise.

In the multi-lingual retrieval setup, we consider a set $\mathbb{L}$ of $n$
languages $\{L_1, L_2, \dots, L_n\}$. Corresponding to each language, there is
a test collection $\mathbb{C}_i = \{\mathbb{D}_i, \mathbb{Q}_i, \mathbb{R}_i\}$
such that documents in $\mathbb{D}_i$ and queries in $\mathbb{Q}_i$ are in
language $\mathbb{L}_i$.  Additionally, the queries in $\mathbb{Q}_i$ are
translations of each other.  In other words, query $q_{kj} \in \mathbb{Q}_k$
(the $j^\text{th}$ query in $\mathbb{Q}_k$) is the translation in language
$\mathbb{L}_k$ of the query $q_{lj} \in \mathbb{Q}_l$ ($l \neq k$) in language
$\mathbb{L}_l$. Each set $\mathbb{Q}_i$ has exactly $m$ queries.

Note that since queries are generally very short phrases and/or just a set of
words, finding translated queries in multiple languages is a much easier task.

\noindent \textbf{Cross-lingual topical relevance hypothesis:} Let
$\mathbb{D}^{R}_{kj}$ denote the set of documents in $\mathbb{D}_k$ that are
relevant to $q_{kj}$. We hypothesize that this set is {\it topically similar}
to the set $\mathbb{D}^{R}_{lj}$ ($l \neq k$) that is relevant to $q_{lj}$
where $q_{kj}$ and  $q_{lj}$ are translations of each other in languages
$\mathbb{L}_k$ and $\mathbb{L}_l$ respectively. Note that the documents in
$\mathbb{D}^{R}_{kj}$ and $\mathbb{D}^{R}_{lj}$ are {\it not} translations of
each other but are supposed to be similar as they are relevant to the
\emph{same} ``information need'' expressed in the two languages $\mathbb{L}_k$
and $\mathbb{L}_l$ respectively.  The sizes of the two sets
$\mathbb{D}^{R}_{kj}$ and $\mathbb{D}^{R}_{lj}$ need not be equal as well.

This notion of topical relevance can be extended to the multi-lingual setting
where for $q_{kj}$'s that are translations of each other, the corresponding
relevant sets $\mathbb{D}^{R}_{kj}$'s are considered topically similar to each
other.

We next describe our proposed method.  We first create a multilingual vector
space embedding of all the $k$ languages together, and then use that to
generate cross-lingual queries that enable retrieval between any two languages
in this multilingual setup.

\begin{figure}[t]
\center
\includegraphics[width=\columnwidth]{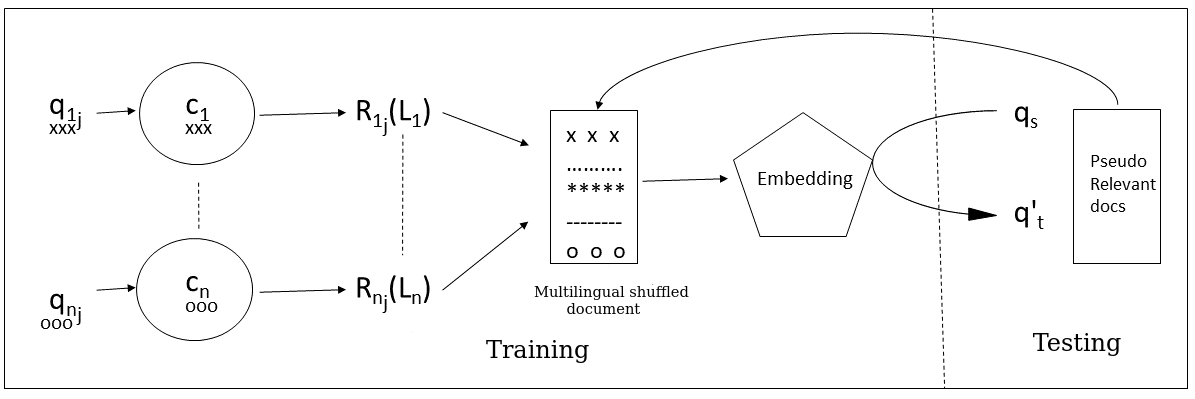}
\vspace*{-8mm}
\caption{Overview of our proposed method.}
\label{fig:overview}
\vspace*{-6mm}
\end{figure}

\subsection{Multilingual Embedding Construction}

In this section, we describe our algorithm for creating a multi-lingual
embedding from all the corpora $\mathbb{C}_i$ designed on the {\it
cross-lingual topical relevance hypothesis} for the set of queries
$\mathbb{Q}_i$. The algorithm is applied for both the training and testing set
of queries.

\noindent \textbf{Training queries:}
We describe the algorithm for creating multi-lingual embedding for one training
query only, which we will generalize for all the training queries thereafter.

Let $q_{ij}$ be the query in language $L_i$ for
which the number of relevant documents is the least among all the corresponding
queries in the other languages $L_k$, i.e., $|\mathbb{D}^{R}_{ij}| \leq
|\mathbb{D}^{R}_{kj}|$ for all $k \neq i$.

For each document $d_i \in \mathbb{D}^{R}_{ij}$, we choose $d_k$ randomly
without replacement from $\mathbb{D}^{R}_{kj}$.  Let $t_{min} = \min \{t_k =
|d_k|\}$ be the minimum document length measured as number of terms among all
the $k$ documents.  Let $d_{min}$ be the corresponding document and $k_{min} =
\arg\min \{t_k\}$ be the language index of $d_{min}$.  Let $n_{k}^{norm} =
\lceil \frac{t_k}{t_{min}} \rceil$.  We create a multi-lingual document
$\mathbb{D}_{j}^{mult}$ comprising of all the $n$ languages as follows.  We
start with an empty $\mathbb{D}_{j}^{mult}$.  For each term in $d_{min}$, we
append the term to $\mathbb{D}_{j}^{mult}$. Then, we select the next
$n_{k}^{norm}$ terms from $d_k$ (for all $k \neq k_{min}$) and append them to
$\mathbb{D}_{j}^{mult}$. Thus, we create $\mathbb{D}_{j}^{mult}$ by placing
each instance of a word of the document which has the least number of terms
followed by the terms of the other documents (of the remaining languages) in
the relative ratios of their document lengths.

This method of shuffling creates a better mix of the words from the multiple
languages, thereby enabling a better learning of the embedded vector space.

For example, let $d_1 = (t_1, t_2)$ and $d_2 = (w_1, w_2, w_3, w_4, w_5)$ be
two documents with terms $t_i$'s and $w_j$'s respectively.  Suppose we are
looking to create a bilingual embedding document from $d_1$ and $d_2$.
Clearly, $t_{min} = 2$.  Then, $n_{2}^{norm} = \lceil \frac{5}{2} \rceil = 3$.
That is, for each term in $d_1$ there will follow $3$ terms of $d_2$ until all
the terms of $d_1$ (and $d_2$) are considered. Therefore,
$\mathbb{D}_{j}^{mult} \equiv (t_1, w_1, w_2, w_3, t_2, w_4, w_5)$.

This algorithm when run for all the training queries produce the set
$\mathbb{D}_{train}^{mult}$.

\noindent \textbf{Test queries:} To address the words that are missing in the
training set of relevant documents (e.g., proper nouns present exclusively in
the test queries), we run the same algorithm by replacing $\mathbb{D}^{R}_{kj}$
with $\mathbb{D}^{PseudoRel}_{kj}$ created by running each test query $q_{kj}$
on the corresponding collection $\mathbb{D}_k$.  We select the top $\kappa$
documents, and run this for all the test queries to produce
$\mathbb{D}_{test}^{mult}$.

\noindent \textbf{Final embedding:} Finally, we create
$\mathbb{D}_{final}^{mult}$ by taking the union of $\mathbb{D}_{train}^{mult}$
and $\mathbb{D}_{test}^{mult}$. We then running Word2Vec \cite{word2vec} on
$\mathbb{D}_{final}^{mult}$ to get the final multi-lingual word vector
embeddings.

\subsection{Cross-Lingual Query Generation}

The main IR task is to perform cross-language information retrieval with a
query $q_{kj}$ in language $\mathbb{L}_k$ on any of the document collections
$\mathbb{D}_l$ in any other language $\mathbb{L}_l$, $l \neq k$. (We exclude
the monolingual setup $l = k$.) The language of the query, $\mathbb{L}_k$, is
referred to as the {\it source language} and the language of the document
collection, $\mathbb{L}_l$, is the {\it target language}.

The aim of cross-lingual query generation is to generate a target query
version, $q'_{lj}$ in language $\mathbb{L}_l$, of $q_{kj}$. Note that this is
different from $q_{lj}$, which is the actual query in language $\mathbb{L}_l$.
In the results section, for reference, we will also state the results using
$q_{lj}$ as the baseline monolingual setting, which is an expected upper-bound
of performance.

\noindent \textbf{Query generation procedure:} Let $\mathbb{V}_l$ be the
\emph{vocabulary} (set of unique terms) of $\mathbb{D}_l$. We construct a
vector $\vec{q}_{kj}$ by aggregating the vectors corresponding to the
constituent terms of $q_{kj}$ in the multi-lingual embedding space.  For each
vector in $q_{kj}$, we capture its top-$\tau$ semantically closest term vectors
from $\mathbb{V}_l$ in the multi-lingual embedding space.  The semantic
closeness is measured by \emph{cosine similarity}.  These closest term vectors
are aggregated to form the target query vector $q'_{lj}$.

Thereafter, we perform cross-lingual retrieval with $q'_{lj}$ on
$\mathbb{D}_l$.

The overall scheme is shown in Figure \ref{fig:overview}.  During training, for
each of the queries, we consider the relevant documents ($R_{kj}$) from the
corresponding corpus, and shuffle them to form a multi-lingual shuffled
document. The multi-lingual document is further enriched by the pseudo-relevant
documents of the test queries. A common word embedding is learned from this set
of multi-lingual documents. During testing, cross-lingual retrieval is done by
generating query $q'$ from the source query $q_s$ using the common word
embedding space.

\section{Experiments}

\begin{table}[t]
 \small
 \begin{tabular}{ccccc}
 \hline
 \multicolumn{4}{|c|}{\textbf{FIRE 2010}}\\
 \hline
 \multicolumn{1}{|c}{\textbf{Language}} & \multicolumn{1}{|c}{\textbf{\#Docs}} & \multicolumn{1}{|c}{\textbf{\#Queries}} & \multicolumn{1}{|c|}{\textbf{Mean rel docs per query}}\\
 \hline
 \multicolumn{1}{|c}{English} & \multicolumn{1}{|c}{1,25,586} & \multicolumn{1}{|c}{50} & \multicolumn{1}{|c|}{13.06}\\
 \multicolumn{1}{|c}{Hindi} & \multicolumn{1}{|c}{1,49,482} & \multicolumn{1}{|c}{50} & \multicolumn{1}{|c|}{18.30}\\
 \multicolumn{1}{|c}{Bangla} & \multicolumn{1}{|c}{1,23,047} & \multicolumn{1}{|c}{50} & \multicolumn{1}{|c|}{10.02}\\
 \hline
 \multicolumn{4}{|c|}{\textbf{FIRE 2011}}\\
 \hline
 \multicolumn{1}{|c}{\textbf{Language}} & \multicolumn{1}{|c}{\textbf{\#Docs}} & \multicolumn{1}{|c}{\textbf{\#Queries}} & \multicolumn{1}{|c|}{\textbf{Mean rel docs per query}}\\
 \hline
 \multicolumn{1}{|c}{English} & \multicolumn{1}{|r}{89,286} & \multicolumn{1}{|c}{50} & \multicolumn{1}{|c|}{55.22}\\
 \multicolumn{1}{|c}{Hindi} & \multicolumn{1}{|c}{3,31,599} & \multicolumn{1}{|c}{50} & \multicolumn{1}{|c|}{57.70}\\
 \multicolumn{1}{|c}{Bangla} & \multicolumn{1}{|c}{3,77,104} & \multicolumn{1}{|c}{50} & \multicolumn{1}{|c|}{55.56}\\
 \hline
 \multicolumn{4}{|c|}{\textbf{FIRE 2012}}\\
 \hline
 \multicolumn{1}{|c}{\textbf{Language}} & \multicolumn{1}{|c}{\textbf{\#Docs}} & \multicolumn{1}{|c}{\textbf{\#Queries}} & \multicolumn{1}{|c|}{\textbf{Mean rel docs per query}}\\
 \hline
 \multicolumn{1}{|c}{English} & \multicolumn{1}{|r}{89,286} & \multicolumn{1}{|c}{50} & \multicolumn{1}{|c|}{70.78}\\
 \multicolumn{1}{|c}{Hindi} & \multicolumn{1}{|c}{3,31,599} & \multicolumn{1}{|c}{50} & \multicolumn{1}{|c|}{46.18}\\
 \multicolumn{1}{|c}{Bangla} & \multicolumn{1}{|c}{3,77,111} & \multicolumn{1}{|c}{50} & \multicolumn{1}{|c|}{51.62}\\
 \hline
 \end{tabular}
 \caption{Datasets.}
 \label{tab:datasets}
 \vspace*{-12mm}
\end{table}

\subsection{Setup}

\noindent \textbf{Datasets:} We use FIRE
(\url{http://fire.irsi.res.in/fire/static/data}) cross-lingual datasets in
English, Hindi and Bangla (details given in Table~\ref{tab:datasets}). The
documents were collected from the following newspapers: The Telegraph
(\url{http://www.telegraphindia.com}) for English, Dainik Jagran
(\url{http://www.jagran.com}) and Amar Ujala (\url{http://www.amarujala.com})
for Hindi, and Anandabazar Patrika (\url{http://www.anandabazar.com}) for
Bangla.  Query sets were created such that queries with the same identifier are
translations of each other. For each language and collection, we choose
randomly $10$ queries for testing. The rest are used for training in a 5-fold
cross validation manner.

For retrieval, only the {\it title} field of the queries were used. Stopword
removal was done. We use the default Dirichlet Language Model implemented in
Terrier IR toolkit (\url{http://terrier.org/}) for all our retrieval
experiments.

\noindent \textbf{Baseline:} We compare our method for cross-lingual IR with
bilingual embeddings with Vulic et al. \cite{Vulic2015}. The shuffling code
used is obtained from the authors.

\noindent \textbf{Mono-lingual:} In the monolingual setup, the results when
the actual target language queries are used for retrieval on the target set
sets the upper bound of performance that can be achieved with a multi-lingual
setup.

\subsection{Training}

The Gensim implementation for Word2Vec
(\url{https://radimrehurek.com/gensim/models/word2vec.html}) was used.  The
skip-gram model was used for the training using the following parameters:
(i)~vector dimensionality: 100, (ii)~learning rate: 0.01, (iii)~min word count:
1. The context window size was varied from 5 to 50 in intervals of 5.  For
bilingual embedding, window size 25 produced the best results on the training
set and was subsequently used on the test queries, while for trilingual
embedding, the best window size was 50.  The parameters $\kappa$ and $\tau$ were
tuned on the training set over the values \{5, 10, 15, 20\} and \{5, 10, 15\}
respectively.


\subsection{Results}

To assess quality, we report the Mean Average Precision (MAP), R-Precision
(R-Prec) and Binary Preference (BPref).

We report our retrieval results in Table \ref{tab:bi} and Table \ref{tab:tri}.
We uniformly use the cross-lingual retrieval convention {\it source language
$\to$ target language}. For example, B$\to$E indicates that Bangla is the
source language while English is the target language.

\noindent \textbf{Bilingual Embeddings:} Table \ref{tab:bi} shows the results
for bilingual retrieval, i.e., when the embedding space is built using only 2
languages.  For all the language pairs, the proposed method outperforms Vulic
et al. \cite{Vulic2015} significantly; the differences are statistically
significant at 5\% level of confidence ($p < 0.05$) by Wilcoxon signed-rank
test~\cite{siegel1956nonparametric}. We have reported the Monolingual results
that does not require any cross-lingual IR as an upper bound of performance.
Interestingly, our proposed method produces comparable MAP results for H$\to$B
(FIRE 2010). It exhibits better BPref than Monolingual B$\to$B for H$\to$B
(FIRE 2010) and the difference is statistically significant at 5\% level of
confidence by Wilcoxon signed-rank test. It is also comparable with Monolingual
H$\to$H for B$\to$H (FIRE 2010), with Monolingual B$\to$B for E$\to$B, H$\to$B
(FIRE 2011) and with Monolingual E$\to$E for H$\to$E (FIRE 2012). While
evaluating with R-Prec, H$\to$B (FIRE 2010) is slightly better than Monolingual
B$\to$B. This shows that the proposed method produces competitive performance
even when compared with a strong baseline like Monolingual.

\noindent \textbf{Time requirements:} The time requirements comparison with
Vulic et al. \cite{Vulic2015} is reported in Table \ref{tab:time}. Our
pre-retrieval time involves indexing time using Terrier
(\url{http://terrier.org}) and cross-lingual query generation time.
Pre-retrieval time for Vulic is the time taken to create the document vectors
for all the documents in a corpus. Retrieval time for us is the one taken by
Terrier to produce the ranked list for only the test queries. Retrieval time
for Vulic comprises of calculating the cosine score between the query vectors
of the test queries and all the documents in the collection followed by sorting
the documents of the whole collection in the decreasing order of this score for
each query. Our proposed method clearly outperforms Vulic in terms of time
requirements.

\noindent \textbf{Trilingual Embeddings:} We report the retrieval performance
of the trilingual setting in Table \ref{tab:tri}.  We chose not to compare with
Vulic et al.  \cite{Vulic2015} any further since we have already established
our superiority over the latter in the bilingual setting. For FIRE 2010, our
proposed method produces superior performance in both MAP and BPref over
Monolingual B$\to$B for both E$\to$B and H$\to$B and the differences are
statistically significant at 5\% level of confidence by Wilcoxon signed-rank
test.  Using R-Prec, for FIRE 2010, E$\to$B is considerably better than
Monolingual B$\to$B ($p < 0.05$ by Wilcoxon signed-rank test). For FIRE 2011,
our proposed method produces better results in BPref over Monolingual B$\to$B
for E$\to$B and over Monolingual H$\to$H for E$\to$H.  This shows that our
proposed method is able to maintain its performance when compared with
Monolingual even in a trilingual setting.

\begin{table}[t]
\small
 \begin{tabular}{|c|c|r|r|}
 \hline
  \textbf{Method} & \textbf{Language} & \textbf{Pre-retrieval time} & \textbf{Retrieval time}\\
  \hline
  Proposed & English & 175.67s & 5.51s \\
  Vulic & English & 10559.37s & 13.37s \\ \hline
  Proposed & Hindi & 2066.27s & 0.92s \\
  Vulic & Hindi & 26206.04s & 36.19s \\ \hline
  Proposed & Bangla & 1627.92s & 3.60s \\
  Vulic & Bangla & 15220.24s & 118.86s \\
  \hline
 \end{tabular}
\caption{Time requirements, averaged over three datasets.}
\label{tab:time}
 \vspace*{-12mm}
\end{table}

\begin{table*}[t]
 \small
 \begin{tabular}{|c|c|c|c|c|c|c|c|c|c|c|}
 \hline
 \multirow{2}{*}{\textbf{Method}} & \multicolumn{4}{|c|}{\textbf{FIRE 2010}} & \multicolumn{3}{|c|}{\textbf{FIRE 2011}} & \multicolumn{3}{|c|}{\textbf{FIRE 2012}}\\
 \cline{2-11}
 & \textbf{Retrieval} & \textbf{MAP} & \textbf{R-Prec} & \textbf{BPref} & \textbf{MAP} & \textbf{R-Prec} & \textbf{BPref} & \textbf{MAP} & \textbf{R-Prec} & \textbf{BPref}\\
 \hline
 \hline
 Monolingual & E$\to$E & 0.4256 & 0.4044 & 0.3785 & 0.2836 & 0.3098 & 0.3528 & 0.4868 & 0.4785 & 0.4507 \\ \hline
Proposed & B$\to$E & 0.1761 & 0.2041 & 0.2297 & 0.1148 & 0.1164 & 0.2204 & 0.2890 & 0.2899 & 0.3418\\
 & H$\to$E & 0.2991 & 0.2548 & 0.2787 & 0.1461 & 0.1810 & 0.2548 & 0.3565 & 0.3861 & 0.4424\\ \hline
Vulic & B$\to$E & 0.0000 & 0.0000 & 0.0000 & 0.0000 & 0.0000 & 0.0000 & 0.0000 & 0.0000 & 0.0000 \\
 & H$\to$E & 0.0000 & 0.0000 & 0.0000 & 0.0000 & 0.0000 & 0.0009 & 0.0000 & 0.0000 & 0.0006 \\ \hline
 \hline
Monolingual & B$\to$B & 0.3354 & 0.2951 & 0.2593 & 0.2127 & 0.2677 & 0.2164 & 0.3093 & 0.3188 & 0.3203\\ \hline
Proposed & E$\to$B & 0.1964 & 0.2429 & 0.2017 & 0.1302 & 0.1797 & 0.2105 & 0.2114 & 0.2409 & 0.2522\\
 & H$\to$B & 0.3108 & 0.3044 & 0.3185 & 0.1098 & 0.1410 & 0.2089 & 0.2058 & 0.2202 & 0.2383\\ \hline
Vulic & E$\to$B & 0.0000 & 0.0000 & 0.0016 & 0.0000 & 0.0000 & 0.0032 & 0.0000 & 0.0000 & 0.0032\\
 & H$\to$B & 0.0001 & 0.0000 & 0.0206 & 0.0000 & 0.0000 & 0.0008 & 0.0000 & 0.0000 & 0.0008\\ \hline
 \hline
Monolingual & H$\to$H & 0.3169 & 0.2872 & 0.2691 & 0.2408 & 0.2756 & 0.2637 & 0.4221 & 0.4407 & 0.4226 \\ \hline
Proposed & E$\to$H & 0.1497 & 0.1663 & 0.1681 & 0.1526 & 0.1806 & 0.2038 & 0.3094 & 0.3093 & 0.3325\\
 & B$\to$H & 0.1791 & 0.2113 & 0.2530 & 0.1244 & 0.1568 & 0.1768 & 0.2751 & 0.2925 & 0.3398 \\ \hline
Vulic & E$\to$H & 0.0000 & 0.0000 & 0.0030 & 0.0000 & 0.0000& 0.0089 & 0.0000 & 0.0000 & 0.0000\\
 & B$\to$H & 0.0000 & 0.0000 & 0.0080 & 0.0000 & 0.0000 & 0.0012 & 0.0000 & 0.0000 & 0.0000\\ \hline
 \end{tabular}
 \caption{Bilingual Retrieval. (Proposed method is always
 statistically significantly better than Vulic et al. \cite{Vulic2015}, $p < 0.05$.)}
 \label{tab:bi}
 \vspace*{-8mm}
\end{table*}

\begin{table*}[t]
	\small
	\begin{tabular}{|c|c|c|c|c|c|c|c|c|c|c|}
		\hline
	\multirow{2}{*}{\textbf{Method}} & \multicolumn{4}{c|}{\textbf{FIRE 2010}} & \multicolumn{3}{c|}{\textbf{FIRE 2011}} & \multicolumn{3}{c|}{\textbf{FIRE 2012}}\\
\cline{2-11}
	& \textbf{Retrieval} & \textbf{MAP} & \textbf{R-Prec} & \textbf{BPref} & \textbf{MAP} & \textbf{R-Prec} & \textbf{BPref} & \textbf{MAP} & \textbf{R-Prec} & \textbf{BPref}\\
	\hline
	\hline
		Monolingual & E$\to$E & 0.4256 & 0.4044 & 0.3785 & 0.2836 & 0.3098 & 0.3528 & 0.4868 & 0.4785 & 0.4507\\ \hline
		Proposed & B$\to$E & 0.2096 & 0.2528 & 0.2243 & 0.1434 & 0.1873 & 0.2497 & 0.3632 & 0.3830 & 0.4370\\
		& H$\to$E & 0.3039 & 0.2981 & 0.2973 & 0.1762 & 0.1887 & 0.2387 & 0.3632 & 0.3854 & 0.4426\\ \hline
		\hline
		Monolingual & B$\to$B & 0.3354 & 0.2951 & 0.2593 & 0.2127 & 0.2677 & 0.2164 & 0.3093 & 0.3188 & 0.3203\\ \hline
		Proposed & E$\to$B & 0.3950 & 0.3651 & 0.3489 & 0.1843 & 0.2062 & 0.2324 & 0.2960 & 0.3198 & 0.3120\\
		& H$\to$B & 0.3558 & 0.2945 & 0.3237 & 0.1566 & 0.1954 & 0.2127 & 0.1941 & 0.2094 & 0.2314\\ \hline
		\hline
		Monolingual & H$\to$H & 0.3169 & 0.2872 & 0.2691 & 0.2408 & 0.2756 & 0.2637 & 0.4221 & 0.4407 & 0.4226\\ \hline
		Proposed & E$\to$H & 0.1759 & 0.2139 & 0.2060 & 0.2259 & 0.2178 & 0.28820 & 0.2774 & 0.2614 & 0.2702\\
		& B$\to$H & 0.1377 & 0.1570 & 0.1833 & 0.1484 & 0.1827 & 0.2160 & 0.2402 & 0.2494 & 0.3082\\ \hline
	\end{tabular}
	\caption{Trilingual Retrieval.}
	\label{tab:tri}
 \vspace*{-8mm}
\end{table*}

\begin{figure*}[t]
	\centering
	\includegraphics[width=0.99\textwidth]{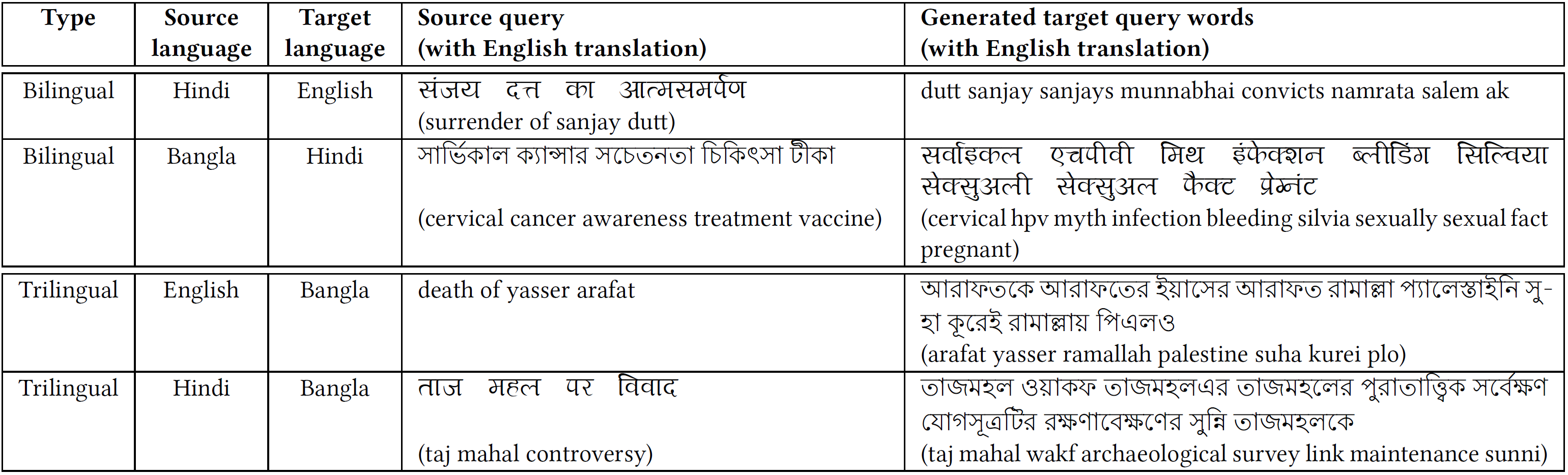}
	\vspace*{-4mm}
	\caption{Target and generated query examples.}
	\label{fig:genquery}
	\vspace*{-6mm}
\end{figure*}

\subsection{Analysis}

Figure \ref{fig:genquery}
shows some example queries generated by our proposed
method using bilingual and trilingual embeddings. For the query {\it surrender
of sanjay dutt} (on the conviction of Bollywood actor Sanjay Dutt with relation
to terrorist attack in Bombay in 1993), the generated query contains important
words such as {\it sanjay}, {\it dutt}, {\it salem} (Abu Salem, a terrorist),
{\it ak} (AK-47, a firearm), {\it munnabhai} (a popular screen name of Sanjay).
The generated query for {\it cervical cancer awareness treatment vaccine}
contains useful terms like {\it cervical}, {\it hpv} (Human papillomavirus),
{\it infection}, {\it pregnant}, {\it silvia} (Silvia De Sanjose, a leading
researcher in Cancer Epidemiology). The generated query for {\it death of
Yasser Arafat} contains the terms {\it arafat}, {\it yasser}, {\it ramallah}
(the headquarters of Yasser), {\it palestine}, {\it suha} (Suha Arafat, Yasser
Arafat's wife) and {\it plo} (Palestine Liberation Organization). The generated
query for {\it taj mahal controversy} (regarding if Taj Mahal is a Waqf
property as claimed by Uttar Pradesh Sunni Wakf Board and subsequent statements
by the Archaeological Survey of India) contains vital terms such as {\it taj},
{\it mahal}, {\it wakf}, {\it archaeological} and {\it sunni}.  These examples
clearly portray the effectiveness of our target query generation.

\section{Conclusions}

In this paper, we have proposed a cross-lingual IR setup in the absence of
aligned comparable corpora.  Our method used a common embedding for all the
languages and produced better performance than the closest state-of-the-art.
In future, we would like to experiment with other embedding methods.

\pagebreak

\balance
\bibliographystyle{ACM-Reference-Format}
\bibliography{arxiv-ref} 

\end{document}